\documentclass[twocolumn,journal,10pt,a4paper,final]{IEEEtran}
\pdfoutput=1
\usepackage[noadjust]{cite}
\usepackage{graphicx,color}
\usepackage[dvipsnames]{xcolor}
\usepackage{amsmath,amsbsy,amsfonts,amssymb}
\usepackage{mathrsfs}
\usepackage{mathtools}
\mathtoolsset{showonlyrefs}
\ifCLASSOPTIONcompsoc
\usepackage[caption=false,font=normalsize,labelfont=sf,textfont=sf]{subfig}
\else
\usepackage[caption=false,font=footnotesize]{subfig}
\fi

\usepackage{acro}
\DeclareAcronym{AWGN}{short = AWGN ,long = additive white gaussian noise}
\DeclareAcronym{AoI}{short = AoI ,long = age of information}
\DeclareAcronym{CDF}{short = CDF ,long = cumulative distribution function}
\DeclareAcronym{CRA}{short = CRA ,long = contention resolution ALOHA}
\DeclareAcronym{CRDSA}{short = CRDSA ,long = contention resolution diversity slotted ALOHA}
\DeclareAcronym{CSA}{short = CSA ,long = coded slotted ALOHA}
\DeclareAcronym{C-RAN}{short = C-RAN ,long = cloud radio access network}
\DeclareAcronym{DAMA}{short = DAMA ,long = demand assigned multiple access}
\DeclareAcronym{DSA}{short = DSA ,long = diversity slotted ALOHA}
\DeclareAcronym{eMBB}{short = eMBB ,long = enhanced mobile broadband}
\DeclareAcronym{FEC}{short = FEC ,long = forward error correction}
\DeclareAcronym{GEO}{short = GEO ,long = geostationary orbit}
\DeclareAcronym{GF}{short = GF ,long = generating function}
\DeclareAcronym{IC}{short = IC ,long = interference cancellation}
\DeclareAcronym{IoT}{short = IoT ,long = internet of things}
\DeclareAcronym{IRSA}{short = IRSA ,long = irregular repetition slotted ALOHA}
\DeclareAcronym{LEO}{short = LEO ,long = low Earth orbit}
\DeclareAcronym{MAC}{short = MAC ,long = medium access}
\DeclareAcronym{mMTC}{short = mMTC ,long = massive machine-type communications}
\DeclareAcronym{MC}{short = MC ,long = Markov chain}
\DeclareAcronym{PDF}{short = PDF ,long = probability density function}
\DeclareAcronym{PER}{short = PER ,long = packet error rate}
\DeclareAcronym{PLR}{short = PLR ,long = packet loss rate}
\DeclareAcronym{PMF}{short = PMF ,long = probability mass function}
\DeclareAcronym{RA}{short = RA ,long = random access}
\DeclareAcronym{rv}{short = r.v. ,long = random variable}
\DeclareAcronym{SA}{short = SA , long = slotted ALOHA}
\DeclareAcronym{SIC}{short = SIC ,long = successive interference cancellation}
\DeclareAcronym{SNR}{short = SNR ,long = signal-to-noise ratio}
\DeclareAcronym{SFG}{short = SFG ,long = signal flow graph}
\DeclareAcronym{TDM}{short = TDM ,long = time division multiplexing}

\begin{document}

\title{\huge On the Value of Retransmissions for Age of Information\\ in Random Access  Networks without Feedback}
\author{Andrea Munari
\vspace{-2em}
\thanks{
The author is with Inst. of Communications and Navigation of the German Aerospace Center (DLR), (e-mail: \mbox{Andrea.Munari@dlr.de}).}
}
\maketitle
\thispagestyle{empty} \setcounter{page}{0}

\begin{abstract}
We focus on a slotted ALOHA system without feedback, in which nodes transmit time-stamped updates to a common gateway. Departing from the classical generate-at-will model, we assume that each transmitter may not always have fresh information to deliver, and tackle the fundamental question of whether sending stale packets can be beneficial from an age-of-information (AoI) standpoint. Leaning on a signal-flow-graph analysis of Markov processes the study reveals that, when packets can be lost due to channel impairments, retransmissions can indeed lower AoI for low generation rates of new information, although at a cost in terms of throughput.
\end{abstract}

\pagestyle{empty}

\newtheorem{prop}{Proposition}
\newtheorem{lemma}{Lemma}

\newcommand{\pr}{\ensuremath{\mathbb P}}
\newcommand{\expOp}{\ensuremath{\mathbb E}}
\newcommand{\de}{\mathrm{d}}

\newcommand{\pmc}{\ensuremath{q}}

\newcommand{\nodes}{\ensuremath{\mathsf n}}
\newcommand{\pGen}{\ensuremath{\alpha}}
\newcommand{\pTx}{\ensuremath{\rho}}
\newcommand{\pTxNew}{\ensuremath{\pi_{\mathsf f}}}
\newcommand{\pTxOld}{\ensuremath{\pi_{\mathsf s}}}
\newcommand{\pTxSame}{\ensuremath{\pi}}
\newcommand{\peras}{\ensuremath{\varepsilon}}

\newcommand{\tru}{\ensuremath{\mathsf S}}
\newcommand{\psucc}{\ensuremath{\mathsf \omega}}
\newcommand{\load}{\ensuremath{\mathsf G}}

\newcommand{\tStamp}{\ensuremath{\tau}}
\newcommand{\age}{\ensuremath{\delta}}
\newcommand{\Age}{\ensuremath{\Delta}}
\newcommand{\overbar}[1]{\mkern 1.5mu\overline{\mkern-3mu#1\mkern-3mu}\mkern 1.5mu}
\newcommand{\avgAge}{\ensuremath{\overbar{\Age}}}
\newcommand{\optAge}{\ensuremath{\avgAge^*}}
\newcommand{\Irt}{\ensuremath{Y}}
\newcommand{\irt}{\ensuremath{y}}
\newcommand{\AStart}{\ensuremath{Z}}
\newcommand{\aStart}{\ensuremath{z}}

\newcommand{\mcR}{\ensuremath{\mathsf R}}
\newcommand{\mcF}{\ensuremath{\mathsf F}}
\newcommand{\mcS}{\ensuremath{\mathsf S}}

\section{Introduction} \label{sec:intro}

\Ac{AoI} has emerged as a key parameter for the design of \ac{IoT} systems where nodes report the status of a monitored process to a central gateway \cite{Yates20_Survey}. Originally introduced in \cite{Kaul11_SECON}, the metric offers a measure of the freshness of the available knowledge, capturing the time elapsed since the generation of the last available update. In spite of its simplicity, \ac{AoI} has been shown to effectively describe the fundamental trade-offs in many machine-type and cyber-physical applications \cite{Uysal20_TIT,Kellerer19}, and an increasing attention has recently been devoted to characterize its behavior in large-scale wireless networks.

In this perspective, first important results have been obtained focusing on grant-free (random) access medium sharing policies \cite{Munari21:Balkancom} which are commonly employed in commercial \ac{IoT} solutions \cite{LoRa}. Assuming a generate-at-will policy, in which a device can collect new information on the process right before sending a packet, the performance of ALOHA \cite{Yates17:AoI_SA,Yates20_ISIT} as well as of more advanced variations of the strategy \cite{Munari21_TCOM_AoI} has been extensively studied. Moreover, significant improvements have been demonstrated in the presence of feedback, allowing each device to adapt its transmission probability to the \ac{AoI}-level \cite{Uysal21_AlohaThresh,Shirin19_ISIT}. In all cases, for slotted ALOHA the access policy maximizing throughput also entails the minimum \ac{AoI}.

Inspired by practical considerations, we depart from these assumptions and consider a slotted ALOHA system in which transmitters may not always have fresh information to deliver. The setting is especially relevant for applications in which an IoT device receives readings from a sensor whenever these become available, and cannot generate fresh data at will. Moreover, we focus on wireless links \emph{without feedback}, commonly encountered in \ac{IoT} networks, e.g. LoRa and Sigfox, to reduce terminal complexity and energy consumption.

In this setup, we tackle the fundamental and open problem of whether sending a packet multiple times can be useful from an \ac{AoI} perspective. In fact, the possibility for a device to perform re-transmissions when fresher information is not yet available triggers a non-trivial trade-off, aiming to increase the possibility for the update to be delivered and reduce \ac{AoI}, yet injecting more traffic onto the random access channel. To characterize this, we focus on three access policies: a plain throughput maximization approach, a reactive strategy in which a device transmits only when it has fresh information, and a retransmission-based policy, where novel and stale packets are sent with different probabilities. We derive closed-form expressions for throughput and average \ac{AoI} for all the solutions, and identify some fundamental insights. Specifically, we show that for a pure collision channel retransmissions are never convenient in terms of \ac{AoI}. Conversely, when packets can be lost due to channel impairments (captured by a simple on-off fading model \cite{Grant03:ISIT}), the transmission of stale information can be beneficial for sufficiently generation rates of fresh data. In this case, we provide simple exact formulations that capture such region, as well as the access probabilities that optimizes \ac{AoI}. Notably, we prove that information freshness can be improved at the expense of throughput, revealing what, to the best of our knowledge, is the first example of such a trade-off in plain slotted ALOHA schemes.

\section{System Model and Preliminaries} \label{sec:sysModel}

We focus on a system in which \nodes\ devices (transmitters) share a wireless channel to send  packets a common destination (gateway). Time is slotted, and the duration of a slot allows transmission of a single packet. Each device receives input from a sensor, which monitors a process. Processes are assumed i.i.d., and a sensor provides to its device a new reading at the beginning of a slot with probability \pGen. The transmitter stores the update until the next one is generated, thereby having at any time a single packet ready for delivery to the gateway,  containing the last produced sensor reading.

Medium access is performed via slotted ALOHA \cite{Roberts72:ALOHA}. Specifically, at any slot a node sends a packet with probability \pTxNew\ if new sensor data has just been generated, whereas the transmission is performed with probability \pTxOld\ whenever the available reading is not fresh. No feedback is provided by the gateway, so that the behavior of a node does not depend on the outcome of previous transmissions.

Throughout our analysis, we consider a collision channel model with erasures.
Accordingly, a sent packet reaches the gateway with probability $1-\peras$, or is completely erased with probability $\peras\geq 0$, bringing no power contribution to the destination.\footnote{In spite of its simplicity, the model has been shown to capture well the fundamental trade-offs of slotted ALOHA channels, see, e.g. \cite{Sun17:TCOM}.}
The number of incoming data units at the gateway over a slot follows thus a binomial distribution of parameters $(\nodes, (1-\peras)\,\pTx)$, where
\begin{align}
  \pTx := \pGen \, \pTxNew + (1-\pGen) \pTxOld
  \label{eq:pTx}
\end{align}
summarizes the probability for a transmitter to access the channel.
Collisions are regarded as destructive, so that no information can be retrieved from slots which see at the receiver the superposition of two or more unerased packets, while data units over singleton slots are always correctly decoded. Under these assumptions, a transmitter accessing the channel successfully delivers its update with probability
\begin{align}
  \psucc = (1-\peras) \left[ 1- \pTx \, (1-\peras) \right]^{\nodes-1}.
\end{align}
Accordingly, the aggregate throughput, i.e. the average number of packets decoded per slot, evaluates to
\begin{align}
  \tru = \nodes \pTx \, \psucc \simeq \nodes\pTx (1-\peras) \, e^{-\nodes\pTx (1-\peras)}
  \label{eq:tru}
\end{align}
where the approximation relies on the well-known limit $\lim\nolimits_{\nodes\rightarrow\infty}(1-a/\nodes)^{\nodes-1} = e^{-a}$ and becomes very tight already for moderately small values of \nodes.
For convenience, we shall refer to the quantity $\nodes\rho(1-\peras)$, describing the average number of non-erased packets incoming at the receiver over a slot, as the \emph{channel load}. Along this line, we recall that throughput is maximized at a load of $1$ [pkt/slot].

\begin{figure}
  \centering
  \includegraphics[width=.75\columnwidth]{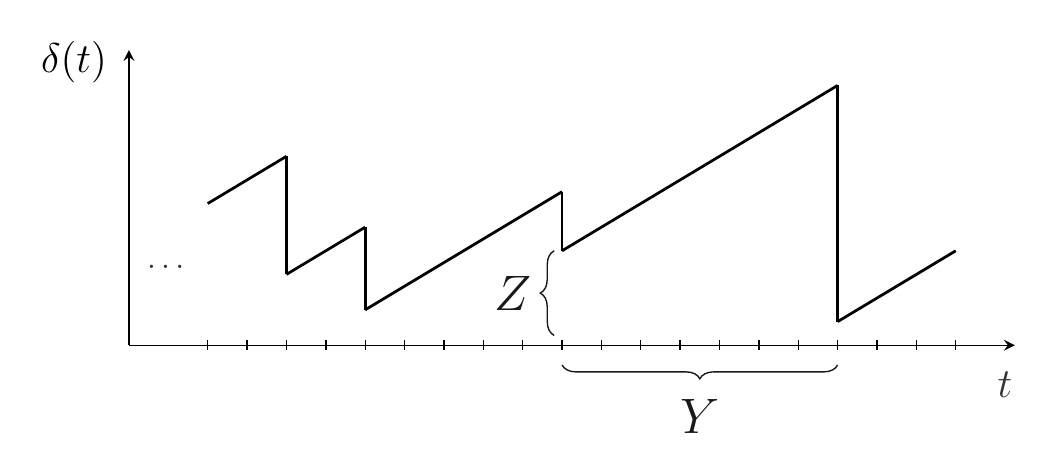}
  \vspace{-1em}
  \caption{Example of a realization $\age(t)$ of the AoI process. The r.v.s for inter-refresh duration \Irt, and current AoI after a refresh, \AStart, are also shown.}
  \label{fig:aoiTimeline}
\end{figure}

In this setting,
we assume each packet to have a time stamp denoting the instant at which the contained information was created, and introduce the current \ac{AoI} for a node at time $t$ as
\begin{align}
  \age(t) := t - \tStamp(t)
\end{align}
where $\tStamp(t)$ is the time stamp of the last update from the sender that was successfully received at the gateway. The corresponding stochastic process is denoted as $\Age(t)$. The metric follows the well-known saw-tooth profile exemplified in Fig.~\ref{fig:aoiTimeline}, growing linearly over time until an \emph{innovative} update from the transmitter of interest is decoded, i.e., a packet with time stamp fresher than what already received at the gateway. In such a case, the age value is reset to the time elapsed since the generation of the update, at least equal to the one slot needed for transmission.  The process $\Age(t)$ is ergodic (see Remark 1), and we will focus in the remainder of our study on its stationary behavior, considering the \emph{average \ac{AoI}} for a device
\begin{align}
  \avgAge := \expOp \left[ \, \Age(t) \, \right].
  \label{eq:avgAoI_def}
\end{align}

\emph{Notation:}
Throughout our discussion, we refer to a \ac{rv} and its realizations with capital and lowercase letters, respectively. For a non-negative discrete \ac{rv} $A$ with alphabet $\mathbb N^0$, we denote the \ac{PMF} as $p_A(a)$, and consider its probability generating function
\begin{align}
G_A(x) := \expOp \!\left[x^A\right] = \sum_{a=0}^\infty p_A(a) \,x^a.
\label{eq:genFunc_def}
\end{align}
From the definition in \eqref{eq:genFunc_def}, the first and second order moment of $A$ readily follow as \cite{Feller}
\begin{align}
  \expOp\left[ A\right] = G_A^{\prime}(1), \quad   \expOp\left[ A^2\right] = G_A^{\prime\prime}(1) + G_A^{\prime}(1)
  \label{eq:moments_genFunc}
\end{align}
where $G_A^{\prime}(x)$ and $G_A^{\prime\prime}(x)$ denote the first and second order derivative of the probability generating function.

Finally, for a finite-state \ac{MC}, we denote the one-step transition probability between states $i$ and $j$ as $\pmc_{i,j}$.

\vspace{.5em}
\emph{Signal Flow Graphs:}
The notion of \ac{SFG}, originally introduced by Shannon in 1942 \cite{Shannon_SFG}, denotes a directed weighted graph in which each vertex $v$ is associated to a variable. The value taken by the variable is in turn given by the sum of the variables of all vertices emitting an edge towards $v$, each multiplied by the corresponding edge weight. As such, a \ac{SFG} offers a simple graphical representation of a system of linear equations, and serves as basis for a number of useful mathematical tools (see, e.g., \cite{Lorens56}). Among them, Mason's gain formula \cite{Mason55} allows to derive the direct dependency between two variables \--- also known as \emph{transfer function} \--- through a simple visual inspection of the graph, i.e., properly identifying paths and cycles connecting the vertices.

In the context of Markov processes, \acp{SFG} provide a convenient approach to compute the moments of recurrence times. In particular, given a finite-state \ac{MC}, an associated \ac{SFG} can be constructed mapping each state into a vertex, and adding a directed edge with weight $x\, \pmc_{i,j}$, where $x$ is a dummy variable, whenever a one-step transition between states $i$ and $j$ is possible. In this case, the transfer function of the \ac{SFG} between a generic vertex $v$ and an absorbing one $w$ (i.e., a vertex with no outgoing edges), corresponds to the probability generating function of the r.v. describing the absorption time for the chain in state $w$ when starting from $v$ \cite{Lorens56}.

\section{Average AoI Analysis}
\label{sec:analysis}

With reference to Fig.~\ref{fig:aoiTimeline}, let us denote by \Irt\ the \emph{inter-refresh} time, i.e., the number of slots that elapse between two successive refreshes of the \ac{AoI} value for a transmitter of interest. It is important to remark that this interval does not generally coincide with the time separating two successive receptions at the gateway of packets from the device. Indeed, the \ac{AoI} is refreshed only upon decoding of a status update containing a time stamp that is more recent than what already available. In other words, receiving retransmitted copies containing the same reading does not improve the perception of the tracked process. Let us further indicate by \AStart\ the r.v. capturing the value of the current \ac{AoI} right after a refresh has taken place.

Following this notation, focus on a generic time $t$, and denote by $\Irt(t)$ the duration of the inter-refresh period the observation instant falls into. Recalling that the \ac{AoI} value grows linearly starting from the value at which it was last reset, under the condition $\AStart=\aStart$, we can write $\Age(t) = z + u$, where $u \in [0,\Irt(t)]$ is the time elapsed between the start of the inter-refresh period and $t$.
From this standpoint, the probability for the observation time to belong to an interval of a specific duration can readily be derived as
\begin{align}
  \pr\{\Irt(t) = \irt \} = \frac{\irt \,p_\Irt(\irt)}{\sum\nolimits_{\irt^\prime} \irt^{\prime} p_\Irt(\irt^{\prime})}.
  \label{eq:pmfY}
\end{align}
Moreover, conditioned on $\Irt(t)$, the observation time is uniformly distributed within the inter-refresh interval. Combining these results, we have
\begin{align}
  \expOp &\left[ \Age(t) \,|\, \AStart \!= \!\aStart \right] = \!\sum_{y=1}^\infty \expOp\left[ \Age(t) \, | \AStart = \aStart, \Irt(t) = \irt \right] \pr\{\Irt(t)=\irt\}\\[-1.5em]
  &=\sum_{\irt=1}^\infty \int_{0}^\irt \frac{\aStart + u}{\irt} d u \cdot \frac{\irt \, p_\Irt(\irt)}{\expOp[\Irt]}
  = \aStart + \sum_{\irt=1}^\infty \frac{\irt^2 p_\Irt(\irt)}{2 \,\expOp[\Irt]}.
  \label{eq:avgAoI_condZ}
\end{align}
Finally, observing that $\Irt$ and \AStart\ are independent, and removing the conditioning on the latter from \eqref{eq:avgAoI_condZ}, a compact expression for the average \ac{AoI} is obtained in the form:
\begin{align}
  \avgAge = \expOp\left[ \AStart \right] + \frac{\expOp [ \Irt^2 ]}{2\, \expOp[\Irt]}.
  \label{eq:avgAoI}
\end{align}

The characterization of $\avgAge$ requires then the calculation of the first and second order moments of some relevant processes.

\emph{Average refresh value: } Let us start by considering the statistics of \AStart, and focus first on a slot over which the \ac{AoI} for the transmitter of interest is refreshed. In this case, the new value of $\Age(t)$ is $1$ if the sender has a fresh update to send, accesses the channel, and the delivery succeeds, i.e., with probability $\pGen\pTxNew\,\psucc$. Conversely, the metric is reset to $k>1$ if and only if three conditions are met: i) a new reading was produced $k$ slots ago and fed to the transmitter, but not immediately delivered to the gateway (probability $\pGen (1-\pTxNew\,\psucc)$); ii) $k-2$ further slots have elapsed without a new readings being generated and without the latest update reaching the gateway (probability $[(1-\pGen)(1-\pTxOld\psucc)]^{k-2}$); and iii) no new reading is acquired in the current slot, but the stale one is transmitted and decoded (probability $(1-\pGen) \pTxOld\, \psucc$). Based on this, the probability to observe a reset of the \ac{AoI} over a time slot, denoted for convenience as $\zeta$, can be computed by simply summing the contributions of the described events, to obtain
\begin{align}
  \zeta &= \pGen\psucc \left( \pTxNew + \sum_{k=2}^\infty \pTxOld (1-\pGen)^{k-1} (1-\pTxNew \,\psucc)(1-\pTxOld\,\psucc)^{k-2}\right)\\
  &= \pGen\psucc \pTx \cdot \left[\pGen + (1-\pGen) \pTxOld\psucc \right]^{-1}.
\end{align}
If we now recall that, by definition, the \ac{PMF} $p_\AStart(\aStart)$ captures the probability for $\Age(t)$ to be reset to $\aStart$ given that a refresh has taken place, we readily obtain
\begin{align}
  p_\AStart(\aStart) \!=\! \frac{\pGen\psucc}{\zeta} \cdot\!
    \begin{dcases}
      \pTxNew & \aStart = 1 \\
      \pTxOld (1-\pGen)^{k-1} (1-\pTxNew \,\psucc)(1-\pTxOld\,\psucc)^{k-2} &\aStart > 1
    \end{dcases}
\end{align}
Accordingly, the first order moment required for the calculation of \eqref{eq:avgAoI} can be derived after simple algebraic manipulations:
\begin{align}
\expOp[\AStart] = \sum_{\aStart=1}^{\infty} \aStart \, p_\AStart(\aStart) = 1 + \frac{\pTxOld(1-\pGen)(1-\pTxNew \psucc)}{\pGen + (1-\pGen) \pTxOld \psucc}.
\label{eq:meanZ}
\end{align}

\begin{figure}[t]
    \centering
    \includegraphics[width=.55\columnwidth]{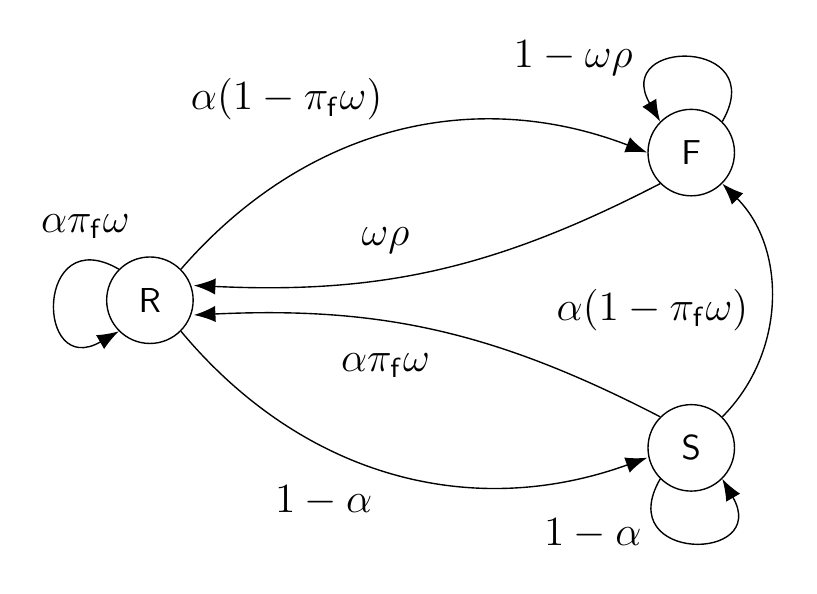}
    \caption{Markov chain to track the behavior of a transmitter. The inter-refresh time corresponds to the recurrence time of state \mcR.}
    \label{fig:MC}
\end{figure}

\emph{Statistical moments of \Irt: }
Let us then focus on the statistics of the inter-refresh time, and consider the \ac{MC} reported in Fig.~\ref{fig:MC}, which usefully characterizes the behavior of a transmitter from an \ac{AoI} perspective. Transitions take place at each slot, across three possible states. Specifically, the chain enters state \mcR\ whenever the transmitter delivers a packet that triggers a reset of the \ac{AoI} value at the gateway. On the other hand, state \mcS\ is visited when the time-stamp of the process reading available at the node matches the one already delivered to the receiver. Note that, in these conditions, a successful packet transmission from the node to the gateway would not lead to an \ac{AoI} refresh. Conversely, the \ac{MC} transitions into state \mcF\ when the transmitter has been fed with fresher information, which would lead to a reset of $\Age(t)$ if sent and successfully decoded.
The transition probabilities among states can readily be computed for the presented system model. For example, the chain will remain in \mcR\ if a new reading is generated, sent and delivered over the slot, i.e. with probability $\pmc_{\mcR,\mcR} = \pGen\pTxNew \psucc$. A jump from \mcR\ to \mcS, instead, takes place with probability $\pmc_{\mcR,\mcS} = 1-\pGen$, corresponding to having no fresher information avaliable at the transmitter compared to the receiver's picture. Similarly, consider state \mcF. From here, the chain moves into \mcR\ whenever an update is delivered from the node, i.e. with probability $\pmc_{\mcF,\mcR} = \pTx \psucc$. Conversely, no jump to state \mcS\ is possible, since the transmitter does have innovative information for the gateway. The remaining transition probabilities are reported for completeness in Fig.~\ref{fig:MC}.

\begin{figure}[t]
    \centering
    \includegraphics[width=.6\columnwidth]{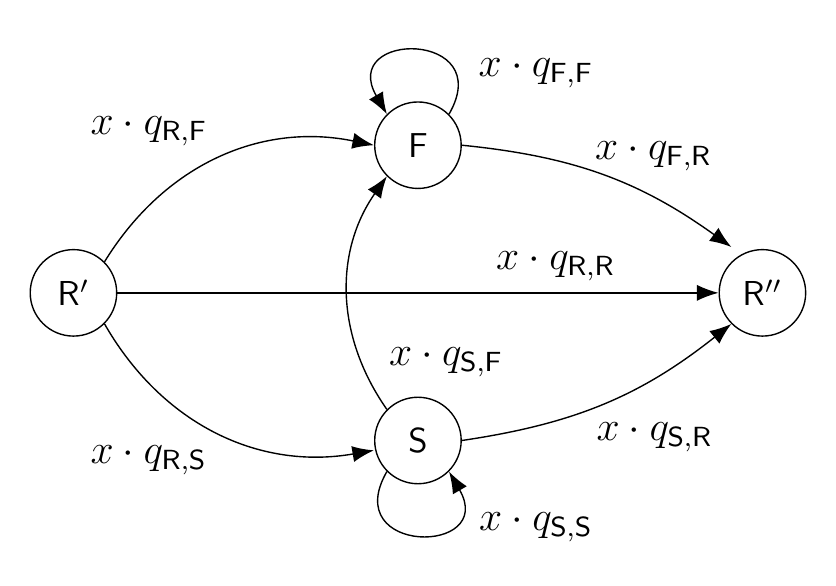}
    \vspace{-.7em}
    \caption{Signal flow graph corresponding to the Markov chain in Fig.~\ref{fig:MC}.}
    \label{fig:SFG}
    \vspace{-1em}
\end{figure}

Leaning on the presented \ac{MC}, the inter-refresh time can be computed as the recurrence time for \mcR, i.e. the number of slots between two successive visits to the state. To this aim, recalling the discussion of Sec.~\ref{sec:sysModel}, we consider the \ac{SFG} associated to the Markov process, reported in Fig.~\ref{fig:SFG}. For convenience, we split \mcR\ into two states: $\mcR^\prime$ (with only outgoing transitions from \mcR) and $\mcR^{\prime\prime}$ (collecting all of \mcR's incoming edges). In this configuration, the transfer function between the two states captures the probability generating function of the absorption time into $\mcR^{\prime\prime}$ when starting from $\mcR^\prime$, which corresponds exactly to the sought inter-refresh period. A direct application of Mason's gain formula \cite{Mason55} provides then the generating function
\begin{align}
  G_\Irt(x) = \frac{\pGen\pTxNew\psucc \cdot x - \pGen\psucc(\pTxNew-\pTx)\cdot x^2}
  {1 - (2-\pGen-\pTx\psucc) \cdot x + (1-\pGen)(1-\pTx\psucc) \cdot x^2 }.
  \label{eq:transFunc}
\end{align}
In turn, the first and second order moments of the r.v. $\Irt$ can be derived computing the derivatives of $G_\Irt(x)$ as highlighted in \eqref{eq:moments_genFunc}, leading after few simple manipulations to the expression
\begin{align}
  \frac{\expOp[\Irt^2]}{2\expOp[\Irt]} = \frac{\nodes}{\tru} + \frac{1}{\pGen} - \frac{1}{2} - \frac{1}{\pGen+(1-\pGen)\pTxOld\psucc}.
  \label{eq:moments_ratio}
\end{align}

Finally, plugging \eqref{eq:meanZ} and \eqref{eq:moments_ratio} into \eqref{eq:avgAoI} leads to a simple closed form formulation of the average \ac{AoI} for a transmitter:
\begin{align}
  \avgAge = \frac{1}{2} + \frac{\nodes}{\tru} + \frac{1}{\pGen} - \frac{\pTxNew}{\pTx}.
  \label{eq:avgAoI_final}
\end{align}

\emph{Remark $1$ (Ergodicity of $\Age(t)$):} Consider the imbedded \ac{MC} capturing the current \ac{AoI} at start of each slot. For this process, the transition probabilities follow from the system model, and the chain can easily be shown to be aperiodic and irreducible. Observe now that at each slot the \ac{AoI} is reset to its minimum value with probability $\pGen\pTxNew\psucc$. The recurrence time for state $1$ of the \ac{MC} is thus geometric distributed, and has mean value $1/\pGen\pTxNew\psucc > 0$. Recalling that this is also the reciprocal of the stationary probability for that state, the chain admits a proper limiting distribution, and is thus ergodic.

\section{Minimum AoI under different Access Policies}

The compact expression of the average \ac{AoI} derived in \eqref{eq:avgAoI_final} conveniently allows to capture the behavior of different transmission strategies. To further delve into this, we tackle in particular three relevant policies, all of which operate without feedback from the gateway.

\emph{Plain throughput optimization:} As a starting point, let us focus on a plain slotted ALOHA scheme, which foresees nodes to access the channel at each slot with probability $\pTxSame$ to send the buffered packet. In other words, transmitters operate agnostically of the freshness of the available reading, using the same probability to perform the first delivery attempt of an update as well as to retransmit it over successive slots until a new one is generated. Note that such an approach represents a benchmark of practical relevance, epitomizing the typical operation of a random access system which does not cope with \ac{AoI}. In this case, $\pTxNew=\pTxOld=\pTxSame$, and $\pTx=\pTxSame$, so that, from \eqref{eq:avgAoI_final}, $\avgAge = \nodes/\tru + 1/\pGen - 1/2$. Accordingly, for any generation rate $\pGen$, the access probability that maximizes throughput also optimizes \avgAge. The minimum \ac{AoI} follows then by setting $\pTxSame = [\nodes(1-\peras)]^{-1}$, obtaining via \eqref{eq:tru}
\begin{align}
  \optAge \simeq \nodes e + \frac{1}{\pGen} - \frac{1}{2}.
  \label{eq:aoi_tru}
\end{align}

\emph{Reactive policy:} A second relevant strategy foresees a node operate in a reactive fashion, accessing the channel with probability $\pTxNew$ only when new information is available. In our setting, this corresponds to setting $\pTxOld=0$, so that $\pTx = \pGen \pTxNew$. This approach ensures that any packet decoded at the gateway is innovative, and triggers a refresh of the \ac{AoI}. On the other hand, in the absence of possible retransmissions, the loss of a message (either due to an erasure or a collision) results in a more stale perception of the monitored process at the receiver. Incidentally, we note that the reactive policy can also be seen as a slotted ALOHA system with generate-at-will traffic model, in which each transmitter attempts delivery of a fresh packet at each slot with probability \pTx. Accordingly, the average \ac{AoI} obtained from \eqref{eq:avgAoI_final} takes the well-known form $\avgAge = \nodes/\tru + 1/2$, see, e.g., \cite{Yates17:AoI_SA,Munari21_TCOM_AoI}.
Also in this case, the optimal behavior in terms of \ac{AoI} coincides then with throughput maximization, obtained by setting $\pTxNew = \min\{1,[\pGen\nodes(1-\peras)]^{-1}\}$ and leading to a minimum average age
\begin{align}
 \optAge \simeq \frac{1}{2} +
 \begin{dcases}
   \nodes e                                                   &\pGen > [\nodes(1-\peras)]^{-1} \\
   \frac{e^{\nodes\pGen(1-\peras)}}{\pGen(1-\peras)}          & \pGen \leq [\nodes(1-\peras)]^{-1}
 \end{dcases}.
 \label{eq:aoi_reactive}
\end{align}

\emph{Retransmission-based policy:} Finally, let us consider a complete policy which allows a transmitter to adapt its access probability, and tackle the fundamental question of whether, in the absence of feedback, transmitting non-fresh information can be useful to reduce \ac{AoI}. To this aim, we start by observing how, (i) for any value of the overall transmission probability $\pTx$, the average age is minimized by picking the maximum possible value of $\pTxNew$.
This is clearly highlighted in \eqref{eq:avgAoI_final}, where such a choice maximizes the absolute value of the \ac{AoI} reduction term $-\pTxNew/\pTx$, and reflects the benefit of being more aggressive in the transmission of fresh rather than stale information. Moreover, \eqref{eq:avgAoI_final} shows that (ii) from an \ac{AoI} perspective it is not convenient to operate the channel in congested conditions, i.e., with average channel load at the receiver larger than $1$ [pkt/slot]. Indeed, it is easy to verify how the derivative of \avgAge\ with respect to \pTx\ is strictly negative for $\nodes\pTx(1-\peras) > 1$, as the negative effect of a drop in throughput more than counterbalances the age reduction brought by the addend  $-\pTxNew/\pTx$.

Combining these two remarks we can conclude that, whenever the overall generation rate of new updates suffices to saturate the channel, i.e., for $\pGen > [\nodes(1-\peras)]^{-1}$, only fresh information shall be sent ($\pTxOld=0$). In this regime the system behaves thus like the reactive policy, and \pTxNew\ shall simply be set to maximize throughput, possibly dropping some newly generated packets.

Consider now the more interesting case \mbox{$\pGen \leq [\nodes(1-\peras)]^{-1}$}. From observation (i), a fresh update shall always be immediately transmitted, setting \mbox{$\pTxNew = 1$}. Accordingly, taking the derivative of the average \ac{AoI} with respect to the overall transmission probability \pTx, we get\footnote{While the presented results lean on the approximation in \eqref{eq:tru}, exact expressions can also be derived considering the binomial formulation of \psucc.}
\begin{align}
  \frac{\partial \avgAge}{\partial \pTx} = \frac{1}{\pTx^2}\left[ 1 + \left(\nodes\pTx - \frac{1}{1-\peras}\right) \, e^{\nodes\pTx(1-\peras)} \right].
\end{align}
For any value of \pGen\ and \peras, the function has a single zero for $\pTx\in [0,1]$, corresponding to a minimum of \avgAge. Specifically, the optimal performance is obtained for an access probability
\begin{align}
  \pTx^*(\peras) = \frac{1 + W\!\left(-(1-\peras) \,e^{-1} \right)}{\nodes(1-\peras)}
  \label{eq:opt_pTx}
\end{align}
where $W(x)$ is the principal value of the Lambert $W$ function, solving in $w$ the equation $w e^w = x \geq 0$. Finally, recalling the composition of \pTx\ reported in \eqref{eq:pTx}, we infer that a node shall send the available stale packet at each slot  with probability  $\pTxOld = \max\{0,(\pTx^*-\pGen)/(1-\pGen)\}$. Note that, having set \mbox{$\pTxNew=1$}, the solution enables in fact retransmissions of the same message, hence the name of the considered policy.

\begin{figure}
  \centering
  \includegraphics[width=.84\columnwidth]{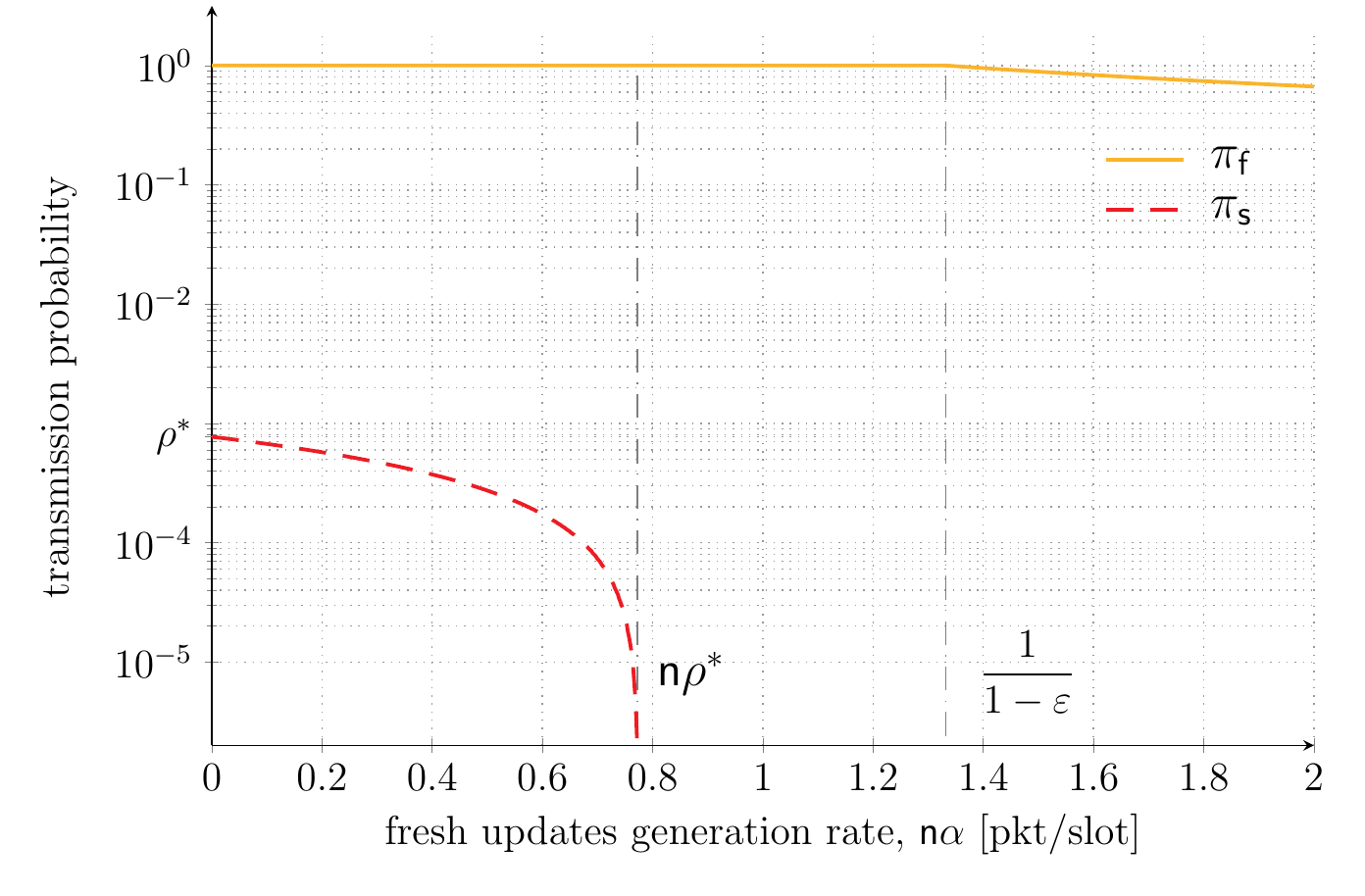}
  \caption{Access probabilities \pTxNew\ and \pTxOld\ minimizing \avgAge\ for the retransmission-based policy, reported against \nodes\pGen. Results obtained for $\nodes=1000$, $\peras=0.25$.}
  \label{fig:optimal_tx}
\end{figure}

Summarizing, the optimal strategy is given as follows:
  \begin{align}
    \begin{split}
    \pTxNew &=
      \begin{dcases}
        \frac{1}{\pGen\nodes(1-\peras)} & \pGen > [\nodes(1-\peras)]^{-1} \\
        \,\,1 & \pGen \leq [\nodes(1-\peras)]^{-1}
      \end{dcases}\\[1em]
      \pTxOld &=
        \begin{dcases}
          \,\,0 & \hspace{1.3em} \pGen \geq \pTx^* \\
          \frac{\pTx^*-\pGen}{1-\pGen} &  \hspace{1.3em} \pGen < \pTx^*
        \end{dcases}
      \end{split}
      \label{eq:optimal_policy}
  \end{align}

\section{Discussion and Results}

The closed form result in \eqref{eq:optimal_policy} is reported graphically in Fig.~\ref{fig:optimal_tx}, and calls for some interesting and non-trivial remarks.

\emph{Observation 1:} In the absence of erasures ($\peras=0$), i.e. for a pure collision channel, retransmissions are never convenient, irrespectively of the generation rate \pGen. Mathematically, the result follows by noting from \eqref{eq:opt_pTx} that $\lim\nolimits_{\peras\rightarrow 0} \pTx^*(\peras) = 0$, so that, from \eqref{eq:optimal_policy}, $\pTxOld = 0$ for any \pGen.
More insightfully, this can be explained focusing on the low channel load region, where retransmissions might play a role. In this regime, the effect of collisions becomes negligible, so that, in the absence of erasures, sent packets will be correctly received. Therefore, for $\peras=0$, the delivery of a fresh update suffices to reset the \ac{AoI} value, and additional transmissions do not bring improvement.

\emph{Observation 2:} Conversely, for any $\peras>0$, sending multiple copies of the same packet in the absence of feedback becomes convenient for sufficiently low generation rates. Specifically, a reduction of the average \ac{AoI} is achieved by properly tuning the access probability \pTxOld\ for any $\pGen \leq \pTx^*$. The traffic threshold for triggering retransmissions, captured by \eqref{eq:opt_pTx}, solely depends on the number of nodes and on the erasure probability. Remarkably, even for very low generation rates ($\pGen\rightarrow 0$), \pTxOld\ shall never exceed a the maximum value of $\pTx^*$. Such an insight is of practical relevance, clarifying how nodes monitoring processes with sporadic changes shall always transmit fresh information, and select instead the channel access probability when stale packets are available based on the $(\nodes,\peras)$ pair.

\emph{Observation 3:} We also note that, for $\pGen \leq \pTx^*$, \ac{AoI} is minimized by operating the system so that the receiver experiences a channel load $\nodes\pTx^* (1-\peras) > 1$. This condition, however, entails a loss in throughput with respect to the maximum achievable performance, obtained for a channel load at the receiver of $1$ [pkt/slot].
This result is particularly interesting, as it pinpoints the existence of a trade-off between throughput and \ac{AoI}. In fact, for low generation rates, maintaining an up-to-date perception of the monitored sources at the gateway comes  at the expense of a less efficient utilization of the channel.

\begin{figure}
  \centering
  \includegraphics[width=.82\columnwidth]{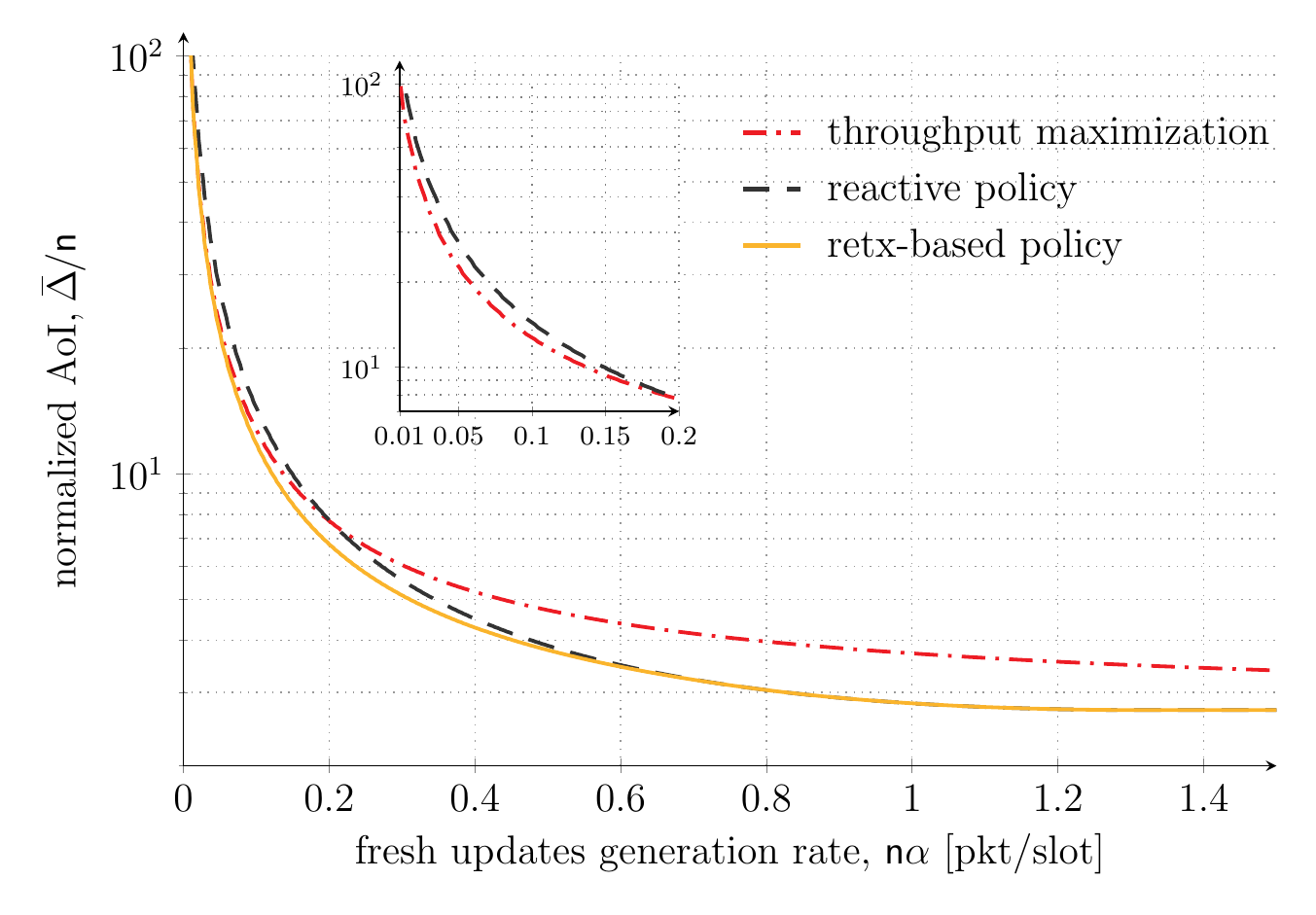}
  \vspace{-.5em}
  \caption{Average AoI normalized to the network population, reported against the average number of new updates generated in the network over a slot ($\nodes\pGen$). Results obtained for $\nodes=1000$, $\peras=0.25$.}
  \label{fig:aoiVsLoad}
  \vspace{-1em}
\end{figure}

To further elaborate on these key trends we report some numerical results, obtained unless otherwise specified in the setting $\nodes=1000$, $\peras=0.25$. As a starting point, Fig.~\ref{fig:aoiVsLoad} shows, for the three considered strategies, the average \ac{AoI} normalized to the network population (i.e., $\avgAge/\nodes$) against the overall average number of new updates generated in a slot (i.e., $\nodes\pGen$). Notably, for high generation rates both the reactive and the retransmission-based policies attain a significant reduction of \ac{AoI} with respect to what achieved via plain throughput optimization. Mathematically, this is confirmed noting that, for  $\pGen > [\nodes(1-\peras)]^{-1}$, the value of \avgAge\ in \eqref{eq:aoi_tru} ($\nodes e +1/\pGen - 1/2$) is always larger than the one of \eqref{eq:aoi_reactive} ($\nodes e + 1/2$), with the two values converging for $\pGen\rightarrow 1$. This result pinpoints the potential of strategies that account for the freshness of available updates, and stems from the fact that the basic throughput optimization policy may
unnecessarily flood the channel with transmissions that would not refresh AoI.

On the other hand, a profoundly different trend emerges when terminals generate new updates less frequently. In this region, Fig.~\ref{fig:aoiVsLoad} reveals that the reactive policy starts to suffer, performing worse than the benchmark throughput maximization approach (see highlight box in the plot). For low values of \pGen, a one-shot transmission of newly generated information may indeed be exceedingly conservative, leading to long update-less periods at the gateway whenever a packet loss is undergone. Under these conditions, the sporadic transmission of stale information may become useful, triggering refreshes of the \ac{AoI} value without the need to wait for new information to be generated. The benefits of the retransmission-based strategy are in turn apparent, as the presented optimization of the channel access probability properly balances the transmission rate based on the availability of fresh updates at each sender.

\begin{figure}
  \centering
  \includegraphics[width=.82\columnwidth]{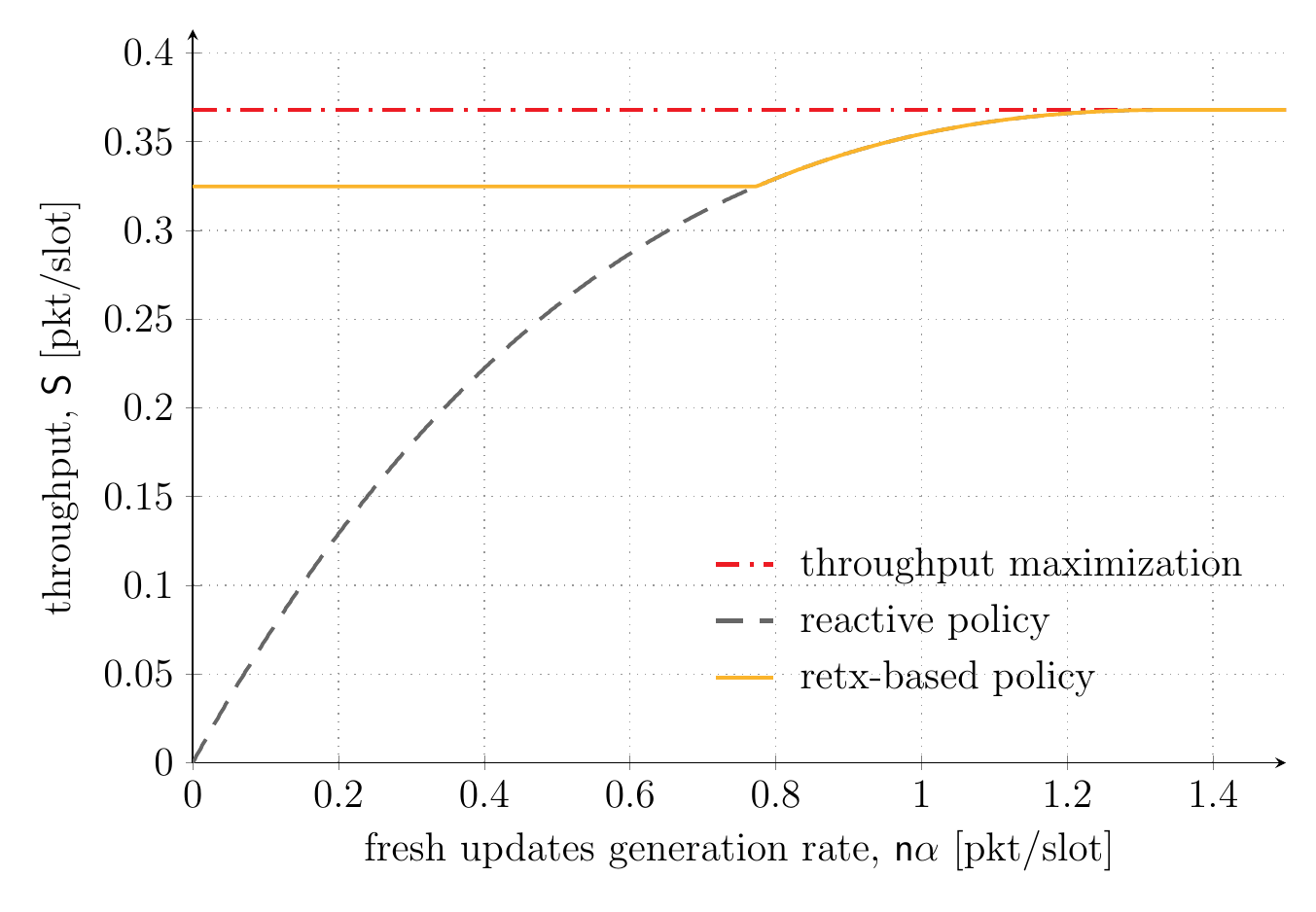}
  \caption{Average throughput vs average number of new updates generated in the network over a slot ($\nodes\pGen$). $\nodes=1000$, $\peras=0.25$.}
  \label{fig:truVsLoad}
  \vspace{-1em}
\end{figure}

As noted earlier, the \ac{AoI} improvements exhibited by the considered strategies entail however a cost in terms of achievable throughput. This fundamental trade-off is explored in Fig.~\ref{fig:truVsLoad}, which reports \tru\ against the overall average number of new updates generated in a slot when the schemes are operated to minimize AoI. By construction, the throughput-optimal policy offers the peak performance of a slotted ALOHA scheme ($\tru \simeq 0.36$ [pkt/slot]) regardless of \pGen\ (dash-dotted line). Conversely, the purely reactive strategy (dashed line) sees a steady throughput reduction as \pGen\ decreases. Note indeed that a more sporadic update generation has the system operate at low channel loads, simply not sharing the channel as efficiently as a slotted ALOHA access could allow throughput-wise.\footnote{This is captured by \eqref{eq:tru}, since for the reactive policy we have \mbox{$\pTxOld=0$}, and, for $\pGen \leq [\nodes(1-\peras)]^{-1}$, $\pTxNew = 1$, so that \mbox{$\tru \simeq \nodes\pGen(1-\peras) e^{-\nodes\pGen(1-\peras)}$}.}
More intestingly, let us focus on the retransmission-based solution. As discussed, for $\pGen > \pTx^*$, the strategy behaves as the reactive one. Instead, for lower generation rates, transmission of non-new updates is possible, tuning the access probability so that the system operates at a load at the receiver of $\nodes\pTx^*(1-\peras)>1$. Recalling the definition of $\pTx^*$ in \eqref{eq:opt_pTx}, the undergone throughput loss can be readily computed taking the ratio of the value of \tru\ attained for this load to the peak value $e^{-1}$ and obtaining:
\begin{align}
  \left[1+W(-(1-\peras) e^{-1})\right]\cdot e^{-W(-(1-\peras)e^{-1})}.
  \label{eq:loss}
\end{align}
Notably, the loss is independent of the number of terminals in the system, and is only driven by the erasure probability \peras. For instance, for $\peras = 0.25$ (see Fig.~\ref{fig:truVsLoad}), the retransmission-based policy attains $\sim 88\%$ of the maximum possible throughput for low \pGen. From \eqref{eq:loss}, the loss becomes even sharper for lower values of \peras, e.g., with half of the peak throughput sacrificed to improve \ac{AoI} for $\peras \sim 0.04$.

\begin{figure}
  \centering
  \includegraphics[width=.84\columnwidth]{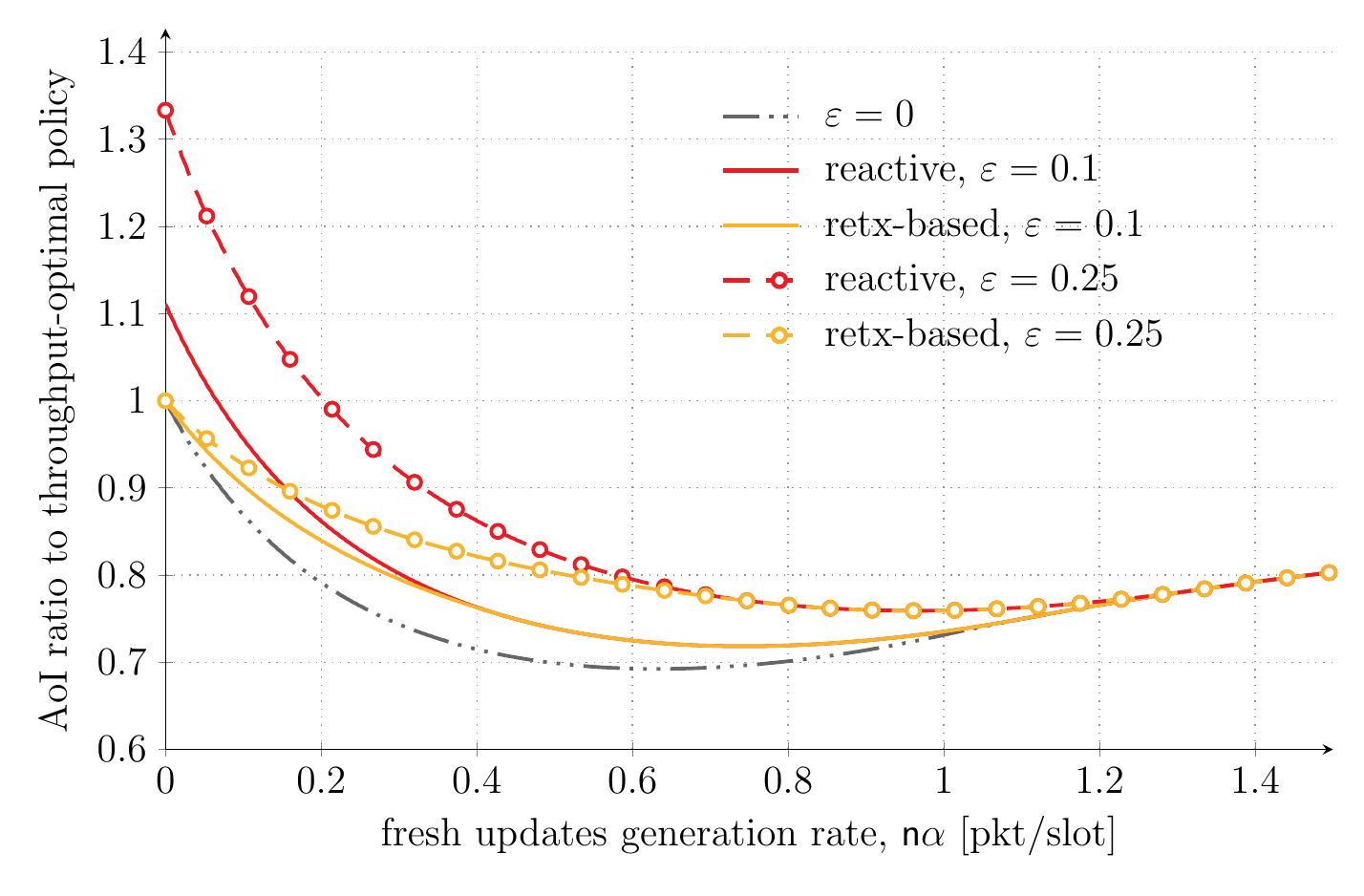}
  \caption{Ratio of the minimum average AoI $\avgAge^*$ for the reactive and retx-based policies to $\avgAge^*$ achieved by the throughput maximizing policy. $\nodes=1000$.}
  \label{fig:aoiRatios}
  \vspace{-1em}
\end{figure}

To conclude, let us explore the impact of the erasure probability. To capture this aspect, we report in Fig.~\ref{fig:aoiRatios} the ratios of $\avgAge^*$ obtained with the reactive and the retransmission-based policies to the one of the throughput maximization approach. The trends are drawn once more against $\nodes\pGen$, and results are shown for different values of \peras. For a pure collision channel ($\peras=0$) a single line is reported (dash-dotted), as both strategies behave in the same way. More relevantly, a reduction of AoI of up to $30\%$ can be obtained by avoiding transmission of stale information. On the other hand, when channel impairments start to play a role, the performance of a purely reactive approach can plummet for low update generation rates. For example, an \ac{AoI} more than $30\%$ higher than the one achieved with a simple throughput optimization approach is experienced for $\peras=0.25$ and low \pGen.
Instead, the retransmission-based approach enables consistent \ac{AoI} reductions under all conditions. From this standpoint, the importance of leaning on, and carefully selecting the access probability for retransmissions clearly emerges.

\section{Conclusions}
In this paper we investigated the role played by retransmissions in terms of age of information in an IoT random access network without feedback. Focusing on a setting in which devices monitoring processes of interest may not always have fresh information to deliver, we considered a practically ALOHA-based channel access and studied three distinct transmission policies. First, a commonly employed throughput optimization approach, which disregards the freshness of packets available at the transmitter side. Second, a reactive solution, foreseeing devices to access the channel only when new information is generated. Finally, we proposed a scheme where a node can adapt its transmission probability based on the availability or not of fresh updates. In all cases, exact closed form expressions for throughput and average AoI were obtained following a signal flow graph analysis of Markov processes. The study revealed that, for a pure collision channel, transmission of stale information is never convenient in terms of AoI. Conversely, when messages may be lost due to channel impairments, allowing a terminal to transmit a packet multiple times (thus attempting delivery of stale information) can indeed be beneficial for sufficiently low generation rates. In such region, a non-trivial trade-off between AoI and throughput emerges.

\bibliographystyle{IEEEtran}
\bibliography{IEEEabrv,biblio_RandomAccess,biblio_AoI}

\end{document}